\documentclass[iicol,pdflatex,sn-mathphys-num]{sn-jnl}

\usepackage{algorithm}
\usepackage{algorithmicx}
\usepackage{algpseudocode}
\usepackage{amsmath,amssymb,amsfonts}
\usepackage{amsthm}
\usepackage{anyfontsize}
\usepackage[title]{appendix}
\usepackage{balance}
\usepackage{booktabs}
\usepackage{enumerate}
\usepackage{float}
\usepackage{stfloats}
\usepackage{graphicx}
\usepackage{listings}
\usepackage{manyfoot}
\usepackage{mathrsfs}
\usepackage{multirow}
\usepackage{textcomp}
\usepackage{xcolor}

\newcommand{\figureref}[1]{Figure~\ref{#1}}

\newcommand{\sectionref}[1]{Section~\ref{#1}}

\newcommand{\tableref}[1]{Table~\ref{#1}}
\newcommand{\eg}{e.\,g.,\ }

\definecolor{red}{RGB}{221, 33, 24}
\definecolor{green}{RGB}{29, 93, 33}
\definecolor{blue}{RGB}{51, 114, 171}
\definecolor{orange}{RGB}{255, 128, 47}
\definecolor{purple}{RGB}{147, 105, 185}

\usepackage[capitalise,nameinlink]{cleveref}
\crefname{section}{Sect.}{Sect.}
\Crefname{section}{Section}{Sections}
\crefname{figure}{Fig.}{Figs.}
\Crefname{figure}{Figure}{Figures}
\crefname{table}{Tab.}{Tabs.}
\Crefname{table}{Table}{Tables}
\crefname{lstlisting}{List.}{List.}
\Crefname{lstlisting}{Listing}{Listings}

\begin{document}

\title[Skyrise: Exploiting Serverless Cloud Infrastructure for Elastic Data Processing]{Skyrise: Exploiting Serverless Cloud Infrastructure \linebreak for Elastic Data Processing}

\author*[1]{\fnm{Thomas} \sur{Bodner}} \email{thomas.bodner@hpi.de}

\author[2]{\fnm{Daniel} \sur{Ritter}} \email{daniel.ritter@sap.com}

\author[1]{\fnm{Martin} \sur{Boissier}} \email{martin.boissier@hpi.de}

\author[1]{\fnm{Tilmann} \sur{Rabl}} \email{tilmann.rabl@hpi.de}

\affil*[1]{\orgdiv{Hasso Plattner Institute}, \orgname{University of Potsdam}, \orgaddress{\city{Potsdam}, \country{Germany}}}

\affil[2]{\orgname{SAP}, \orgaddress{\city{Walldorf}, \country{Germany}}}

\abstract{Serverless computing offers elasticity unmatched by conventional server-based cloud infrastructure. Although modern data processing systems embrace serverless storage, such as Amazon S3, they continue to manage their compute resources as servers. This is challenging for unpredictable workloads, leaving clusters often underutilized. Recent research shows the potential of serverless compute resources, such as cloud functions, for elastic data processing, but also sees limitations in performance robustness and cost efficiency for long running workloads. These challenges require holistic approaches across the system stack. However, to the best of our knowledge, there is no end-to-end data processing system built entirely on serverless infrastructure. In this paper, we present Skyrise, our effort towards building the first fully serverless SQL query processor. Skyrise exploits the elasticity of its underlying infrastructure, while alleviating the inherent limitations with a number of adaptive and cost-aware techniques. We show that both Skyrise's performance and cost are competitive to other cloud data systems for terabyte-scale queries of the analytical TPC-H benchmark.}


\maketitle

\section{Introduction}
\label{section:introduction}

Organizations increasingly move their data processing workloads to the cloud. One main driver is elasticity, which allows them to deploy virtual resources on demand and avoid the upfront cost of on-premise deployment. In particular, data warehousing and analytics with high and fluctuating resource demands benefit from flexible provisioning.
The revenue of cloud-based systems, such as Redshift \cite{Gupta:2015}, BigQuery \cite{Melnik:2010}, and Snowflake \cite{Dageville:2016} lately surpassed the revenue of on-premise systems  \cite{dbms-market-2021} and continues to grow rapidly \cite{dbms-market-2023}.

In the last decade, public cloud infrastructure has evolved to become more elastic. This evolution includes scalable datacenter networking \cite{Singh:2015, Firestone:2018}, accelerated server virtualization \cite{nitro}, and finer-grained economical models for compute resources \cite{ec2-pricing-fine, gcp-pricing}. The benefits of these features, however, do not translate to cloud-hosted systems automatically and require architectural changes.

Systems are being rearchitected to exploit the elasticity on the cloud infrastructure level \cite{Armenatzoglou:2022, Melnik:2020, Vuppalapati:2020}. There are three major design changes: First, cloud systems disaggregate their compute and storage resources to scale them independently. Second, they split their work into small and idempotent tasks that can be scheduled quickly, as well as redundantly in the case of transient errors in the infrastructure. Third, they optimize cost given the economics of their infrastructural building blocks.

Serverless cloud computing is a continuation of this evolution, with most of the related challenges intensifying. Serverless compute units are tiny and stateless \cite{lambda, maf}. Large-scale data systems require many of them and, as a result, are prone to stragglers \cite{Jonas:2017, Ustiugov:2021}. They need to decompose stateful and stateless components with greater care, due to the inefficiencies of remote storage access \cite{Seemakhupt:2023, parquet}. The unit prices are substantially higher, because providers provision serverless resources on behalf of the users to provide elasticity \cite{aws-serverless, gcp-serverless}.

Previous work shows the viability of systems built on serverless infrastructure for infrequent and interactive analytical workloads \cite{Mueller:2020, Perron:2020}. These vary over time and cannot be accurately predicted. One example is a few-shot, resource-intensive query on cold data. This workload category is also targeted by serverless databases that integrate elasticity at the system level \cite{athena, redshift-serverless, snowflake-serverless}.
 
In this paper, we present Skyrise, the first fully serverless query processor. Skyrise has no server-based component and scales down to zero when there is no load. Skyrise offers techniques to improve performance robustness and cost efficiency.
We make the following contributions:
\begin{enumerate}[(1)]
    \item We present the Skyrise query processor, which is an open-source end-to-end SQL system that builds entirely on serverless infrastructure.\footnote{Skyrise is available at \url{https://github.com/hpides/skyrise}.}
    \item We propose adaptive techniques for robust query performance on performance-variable cloud resources. Specifically, we describe the re-triggering of straggling query workers.
    \item We introduce a cache for intermediate results backed by serverless storage, making Skyrise more cost-efficient for high query volumes.
    \item We evaluate the performance, cost efficiency, and elasticity of Skyrise. We show that runtime and cost for TPC-H queries are on par with that of other cloud data systems.
\end{enumerate}

The rest of the paper is structured as follows. In \sectionref{section:serverless_infrastructure}, we review background on serverless cloud infrastructure. Then, we give an overview of the architecture of Skyrise and explain key design decisions in \sectionref{section:system_architecture}. We then evaluate the performance, cost efficiency, and elasticity of Skyrise in \sectionref{section:experimental_evaluation}. We discuss related work in \sectionref{section:related_work} and conclude in \sectionref{section:conclusion}.

\section{Serverless Infrastructure}
\label{section:serverless_infrastructure}

In recent years, serverless cloud infrastructure has matured into a set of reliable services which are effectively utilized in production settings. All major cloud providers have serverless compute (\eg Azure Functions \cite{maf} and AWS Lambda \cite{lambda}) and storage offerings (\eg Amazon S3 \cite{s3} and Google Cloud Storage \cite{gcp-storage}).

The services provide elastic and automatic scalability through large multi-tenant pools of resources.
Besides elasticity and resource management covered by cloud providers, billing of resources at fine consumption granularities allows for a pay-per-use (PPU) economic model.
For that, providers need to match the demand closely and instantly to avoid provisioning/cost of idle capacity, and consequently provision resources on behalf of the users, which result in higher unit prices.
To protect the service, quotas are enforced per user through admission control.

Subsequently, we discuss the two types of infrastructure that resemble core building blocks for data processing systems, namely function as a service (FaaS) platforms for computation and serverless storage services.

\subsection{Function as a Service Platforms}

Function as a service (FaaS) platforms, such as AWS Lambda are realized by combining client-facing load balancing, cluster management, and functions for which the system communicates through eventing.
The cloud function user submits application binaries to be executed as ZIP archives or container images through a frontend service, coordinating function invocations.
The request frontend checks the function metadata with the admission (control) service of the cluster management, and---if the request does not exceed the user's quota of concurrent function executions---it issues sandboxed environments from a pool of hosts for the execution of the binary.

\begin{table}[bt]
\centering
\caption{Sizing and pricing of AWS compute services.}
\vspace{-0.3cm}
\setlength\tabcolsep{3.8pt}
\renewcommand{\arraystretch}{0.8}
\begin{tabular}{lrr}
\toprule
\textbf{Resource} & \textbf{Lambda} (ARM) &\textbf{EC2} (C6g) \\
\midrule
\textbf{Memory} & Configurable & Configurable \\
Capacity [GiB] & 0.125 -- 10 & 2 -- 128 \\
Price [\textcent/GiB-h] & 3.84 -- 4.80 & 0.65 -- 1.70 \\
\midrule
\textbf{Compute} & Memory-based  & Configurable \\
Capacity [vCPU]
& 0.07 --  5.79 & 1 -- 64 \\
Price [\textcent/vCPU-h] & 6.79 -- 8.49 & 1.30 -- 3.40 \\
\midrule
\textbf{Network} & Constant & Compute-based \\
Bandwidth [Gbps] & 0.63 & 0.375 -- 25 \\
Price [\textcent/Gbps-h] & 0.48 -- 0.60 & 3.27 -- 8.70 \\
\bottomrule
\end{tabular}
\label{table:2_compute_services}
\vspace{-0.3cm}
\end{table}

Current FaaS platforms limit function configuration as shown in \tableref{table:2_compute_services}. Sizing is based on memory capacity, which determines the number of virtual CPUs \cite{lambda-limits}.
Notably, function sizes are an order of magnitude smaller than VMs, such as EC2 instances \cite{ec2-c6g}. Functions are short-lived and cannot maintain state beyond their lifetime. They can only communicate indirectly through cloud storage. In addition, they are prone to stragglers, e.g., during initialization or when interacting with remote storage.

For elastic scaling, serverless compute requires quick startup times of hundreds or even thousands of serverless functions.
In \tableref{table:2_compute_startup}, we depict the different startup times of a serverless function that was either freshly created and invoked (cold start) or created, invoked, and invoked again (warm).
Cold starts show 20 to 50 times higher initialization latencies than warm starts, also with higher variation, which can be explained by the clean-up strategy of unused underlying virtual machines.

\subsection{Serverless Storage Services}

Cloud storage suitable for serverless compute requires elastically scalable capacity with high availability and durability to serve concurrent requests from a high number of serverless workers.
The canonical choices are object stores (\eg Amazon S3 \cite{s3}, Azure Blob Storage \cite{mabs}), key-value stores (\eg DynamoDB \cite{ddb-serverless}, GCP Firestore \cite{gcp-firestore}) or shared filesystems (\eg EFS \cite{efs-serverless}, Azure Files \cite{mafiles}).
Object storage is designed to store immutable binary objects of varying sizes and formats at cloud network bandwidth.
In contrast, Key-value stores and networked NFS/SMB file systems have lower latency key lookups at higher IOPS for kilobyte-sized values, but lower bandwidth and throughput.

Current storage infrastructures utilize similar architecture components for load balancing and rate limiting as serverless compute.
The interaction is done via HTTPS requests that are evenly distributed across storage servers that hold partitioned data as shards.
The economic model of cloud storage features static storage costs and dynamic costs per usage for requests and transfers, thus fitting well to the serverless PPU model.

\begin{table}[bt]
\centering
\caption{Startup latency of AWS compute instances.}
\vspace{-0.3cm}
\setlength\tabcolsep{8.2pt}
\renewcommand{\arraystretch}{0.8}
\begin{tabular}{lrrr}
\toprule
\textbf{Latency} [ms] & \textbf{Min} & \textbf{Max} & \textbf{Avg} \\
\midrule
Lambda Cold Start & 122 & 451 & 185 \\
Lambda Warm Start & 5 & 9 & 6 \\
EC2 Cold Start & 12,795 & 22,817 & 15,226 \\
EC2 Warm Start & 9,810 & 19,288  & 11,512 \\
\bottomrule
\end{tabular}
\label{table:2_compute_startup}
\vspace{-0.3cm}
\end{table}

Disaggregated cloud storage exhibits a latency orders of magnitude higher than local storage, as shown in \tableref{table:2_storage_services} for median and tail latencies.
For one million 1 KiB read and write requests, S3 Standard shows the highest latencies of $27$ms read and $40$ms write latency.
S3 Express trades availability for performance and achieves latencies comparable to DynamoDB.

\begin{table*}[bt]
\fontsize{8pt}{8pt}\selectfont
\centering
\caption{Performance and cost characteristics of AWS serverless storage services.}
\setlength\tabcolsep{10.0pt}
\begin{tabular}{lrrrrrrrrr}
\toprule
\textbf{Service} & \multicolumn{2}{r}{\textbf{Median Latency}} & \multicolumn{2}{r}{\textbf{Tail Latency}} & \multicolumn{2}{r}{\textbf{Requests}} & \multicolumn{2}{r}{\textbf{Transfers}} & \textbf{Storage} \\
 & \multicolumn{2}{r}{[ms]} & \multicolumn{2}{r}{[ms]} & \multicolumn{2}{r}{[\textcent/M]} & \multicolumn{2}{r}{[\textcent/GiB]} & [\textcent/GiB-mo] \\
 & Read & Write & Read & Write & Read & Write & Read & Write & \\
\midrule
S3 Standard & 27 & 40 & $>$1k & 500 & 40 & 500 & 0 & 0 & 2.1--2.3\\
S3 Express & 5 & 8 & 120 & 150 & 20 & 250 & 0.15 & 0.8 & 16\\
DynamoDB & 4 & 6 & 100 & 250 & 25 & 125 & 0 & 0 & 25\\
EFS & 6 & 15 & 100 & 600 & 0 & 0 & 3 & 6 & 16--30\\
\bottomrule
\end{tabular}
\label{table:2_storage_services}
\end{table*}

From the cost perspective, S3 Standard is the most economical storage choice with the lowest static storage costs, no transfer costs, but the highest costs for the number of requests.
In contrast, S3 Express halves the request costs, but introduces transfer costs and seven times higher storage costs.
The other storage options have around 10$\times$ higher storage costs compared to S3 Standard and either demand costs per request for DynamoDB or data transfers in case of EFS.

\section{System Architecture}
\label{section:system_architecture}

In this section, we introduce the Skyrise serverless query processor \cite{Bodner:2020}. We provide an overview of its architecture and describe the core components, which fall into three layers: Query compilation, query execution, and storage management. 

Skyrise employs a distributed shared-storage architecture based on serverless object storage. Data referenced by queries always originate and materialize on object storage. The Skyrise query coordinator and worker nodes run as serverless functions. They run only when actively involved in query processing. This way, Skyrise fully exploits the elasticity of modern cloud infrastructure.

Skyrise is written in {\raise.17ex\hbox{$\scriptstyle\mathtt{\sim}$}}27K lines of C++ code (8.5K for query compilation, 10.5K for execution, and 8K for storage management). We adopt the SQL frontend, expression and type systems from Hyrise \cite{Dreseler:2019} and integrate the AWS infrastructure services: Lambda for compute, SQS for messaging, as well as S3, DynamoDB, and Glue for storage. We choose the AWS cloud environment, because it is the most widely used \cite{datadog-report} and studied \cite{Scheuner:2020}.

\subsection{General Overview}

\begin{figure}[bt]
    \centering
    \includegraphics[width=1.0\linewidth]{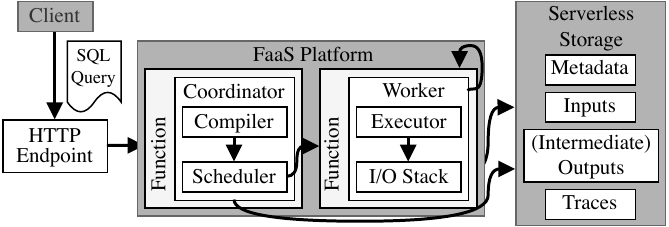}
    \vspace{-0.3cm}
    \caption{Architecture of the Skyrise query processor showing the component interactions for processing a query.}
    \vspace{-0.2cm}
    \label{figure:skyrise_query_processor}
\end{figure}

The main components of Skyrise and their interactions during query processing are depicted in \figureref{figure:skyrise_query_processor}. To process a query, a user sends a SQL query string wrapped inside a JSON request to an HTTP endpoint \cite{lambda-urls}. This triggers a serverless function running an instance of the Skyrise query coordinator. The coordinator gets the query string and manages the lifecycle of just this single query. Additional calls to the same endpoint do not lead to queries being queued. Instead, the queries are managed by separate coordinator instances and run concurrently. The coordinator then parses the SQL string and compiles an executable query plan (cf. \sectionref{section:query_compilation}). This plan contains pipelines of physical operators and dependencies between the pipelines. For each pipeline, the plan defines the number of fragments for parallel execution. Then, the coordinator schedules the pipelines for staged execution based on their dependencies (see \sectionref{section:query_execution}). When a pipeline starts, a serverless query worker function is invoked for each fragment. The worker instances begin local execution by reading their data partitions from shared storage (\sectionref{section:storage_management}). Then, they process their query fragments and write the results back on shared storage. Once all workers finish, the pipeline ends and subsequent pipelines are scheduled until the query completes.

\subsection{Query Compilation}
\label{section:query_compilation}

\begin{figure}[bt]
    \centering
    \includegraphics[width=1.0\linewidth]{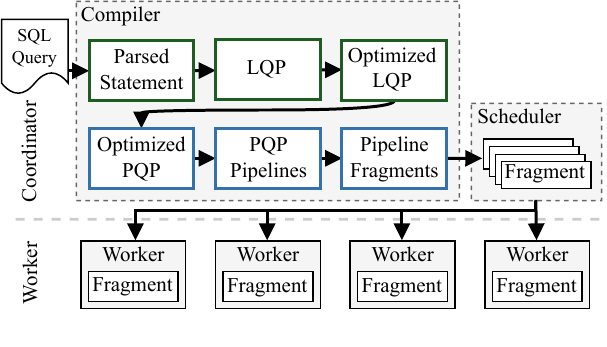}
    \vspace{-0.7cm}
    \caption{Query compilation pipeline showing the intermediate and final compilation artifacts and their consumption.}
    \label{figure:query_compilation}
\end{figure}

\begin{figure*}[ht]
    \centering
    \includegraphics[width=\textwidth]{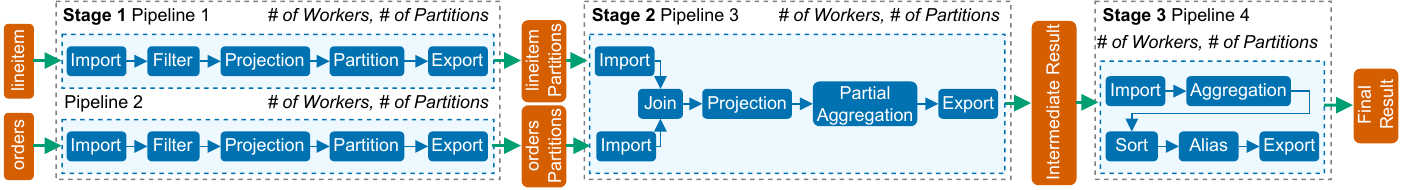}
    \vspace{-0.3cm}
    \caption{Query execution plan for TPC-H Q12.}
    \label{figure:tpch_q12_plan}
\end{figure*}

The query compilation pipeline in Skyrise takes a SQL query and generates a physical plan for distributed data-parallel execution by serverless query workers. The pipeline consists of multiple steps and is depicted in \figureref{figure:query_compilation}. The first set of steps resembles the SQL frontend (in \textcolor{green}{green}). This component parses a query string and builds an abstract syntax tree (AST) that is further transformed into a logical plan for this query (LQP). Thereby, semantic types of columns in referenced tables are validated against an external database catalog. Then, a rule-based optimizer framework applies a range of conventional heuristics-based rules to the plan structure \cite{Dreseler:2020}. These rules include predicate pushdown, subquery flattening, and join ordering. They only need simple statistics and are oblivious of the distributed serverless execution environment of Skyrise. Next, a physical optimizer (in \textcolor{blue}{blue}) employs another set of rules to produce a physical query plan (PQP). The rules map logical operators to physical operators, e.g., a join to a repartition vs. broadcast join. They identify query pipeline breakers in the plan to introduce shuffle points as necessary. And finally, they prepare the plan for serverless execution. The number of workers per pipeline is based on the total input size and the network burst capacity per function, as determined in prior work \cite{Bodner:2025a}. The strategy for shuffling takes the request rate limits of object storage into account and reduces runtime via tiering to hotter storage \cite{s3-eoz}. Since function workers are stateless, the optimizer does not consider cache affinity, and thus data locality. Moreover, the optimizer does not distinguish cold vs. warm starts of functions, because cold starts are negligible (cf. \tableref{table:2_compute_startup}) and only occur in the initial query stage. Subsequent stages reuse the cached function binaries. The final artifacts of query compilation are the per-pipeline PQPs parameterized with the input and output partitions for the individual workers. An example query plan with its pipelines is shown in \figureref{figure:tpch_q12_plan}.

\subsection{Query Execution}
\label{section:query_execution}

The execution of a query in Skyrise is orchestrated by the query coordinator (cf. \figureref{figure:skyrise_query_processor}). Once a query is compiled into a set of interdependent pipeline PQPs, the coordinator schedules these PQPs stage-wise based on their dependencies. For each fragment of a pipeline, the coordinator invokes a serverless function running an instance of the Skyrise query worker. When the number of fragments is large and function invocation can dominate query runtime, the coordinator employs a two-level invocation procedure \cite{Mueller:2020} to parallelize the cluster startup. For $W$ fragments, it invokes $\sqrt{W}$ workers, each with a list of $\sqrt{W}$ fragments that in turn invoke $\sqrt{W}-1$ workers before executing their own fragment. The workers and the coordinator both invoke cloud functions asynchronously, decoupling their lifetime from the lifetime of their direct child functions. Functions are neither invoked speculatively before query arrival nor are they kept running between stages to avoid the cost of idling resources. A function request is a query fragment serialized in JSON. Upon startup, a query worker first deserializes its invocation payload into the PQP form. Then, it schedules and executes the operators accordingly. The worker fetches its input data partitions from shared storage (cf. \sectionref{section:storage_management}) and transforms them into an in-memory columnar format. Since worker functions cannot persist data locally, this process is repeated for recurring queries. From there, the execution follows a push-based, vectorized model. The query operators push their results in batches to downstream operators. Operators with no interdependencies run in parallel. Operators further exploit the thread-level parallelism available in serverless function workers. The final operator in a fragment materializes the results to shared storage. Before the worker quits, it sends a response message to a queue that is polled by the coordinator. The response contains the result location and execution statistics that can be used for adaptive execution behavior and query billing. When the coordinator has received responses from all workers in a stage, it proceeds with scheduling the next stage until the query finishes. Upon query completion, the coordinator responds to the user with the location of the final query result. Skyrise offers robust query performance on its fine-grained resources, i.e., many small tasks and requests, that may fail individually. To accomplish this, Skyrise employs a variety of adaptive techniques. Skyrise's coordinator tracks the progress of workers, which can encounter failures, e.g., because of code issues, data skew, or transient cloud  infrastructure errors. Based on query context and runtime statistics, the coordinator differentiates between the failure categories. Then, the coordinator restarts individual workers, reassigns fragments to more workers, or aborts the query. Retriggering workers mid-query is practically viable, because it takes milliseconds. It does not invalidate (intermediate) query results, because the worker function is idempotent and only ever writes a single deterministic output file to shared storage. Racing worker functions merely overwrite their (identical) results. Aborted queries can be continued from any complete stage result, since they resemble checkpoints. Query failures can be also investigated post-mortem \cite{Mahling:2023}.

\subsection{Storage Management}
\label{section:storage_management}

\begin{figure}[bt]
    \centering
    \includegraphics[width=1.0\linewidth]{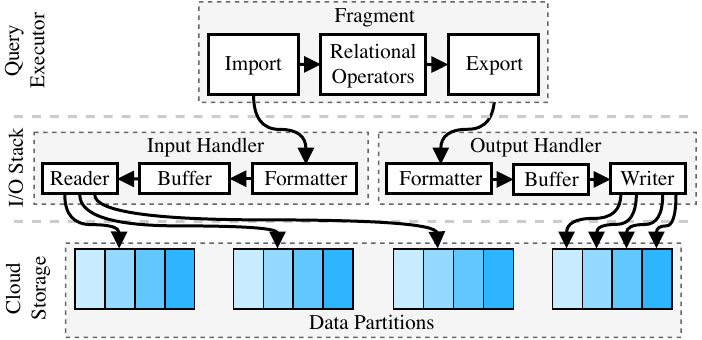}
    \vspace{-0.3cm}
    \caption{Storage stack of the Skyrise query worker including the interactions between query operators and cloud storage.}
    \vspace{-0.3cm}
    \label{figure:storage_management}
\end{figure}

Skyrise disaggregates storage from compute, and as a result, it is subject to the inefficiencies around remote storage access \cite{Seemakhupt:2023, Durner:2023}. In this section, we describe the Skyrise storage stack, which employs a number of techniques to mitigate these issues.

Skyrise uses serverless object storage \cite{s3, gcp-storage} for reading and writing table data during query execution. Thus, its storage stack is designed to efficiently operate on data stored in large objects and formatted in PAX-style \cite{Ailamaki:2002} file formats, such as Parquet \cite{parquet} and ORC \cite{orc}. The key techniques in this setting are right-sizing and parallelizing storage requests, as well as deeply integrating with the query execution and data format components.

\figureref{figure:storage_management} shows the components in the Skyrise storage layer and how they interact with the query execution layer and external cloud storage. There are two dedicated handlers for storage I/O operations that are decoupled from the core query execution and backed by a separate thread pool. This separation accounts for the different resource demands of the two layers.

The input handler splits large object storage requests from the execution layer into several smaller requests, which are processed in parallel. The byte offsets of the small requests are chosen based on the object's PAX layout. In this way, only relevant columns and rows are fetched. The result of each request is stored in a dedicated buffer. Straggling requests are retriggered aggressively after a short timeout. Once buffered, the results are deserialized and decompressed batchwise and on demand, keeping the memory footprint of the I/O stack small. The output handler takes data batches from the execution layer and serializes, compresses, and buffers them as they arrive. Once the query worker's result is complete, it is written to shared cloud storage as a single object.

All objects resulting from a query pipeline combined represent a (intermediate) query result. Query results are referenced in a central registry with an identifier that is generated during query compilation. In particular, the identifier is a hash on the PQP after logical optimization, yet before most of the physical optimization. This allows matching semantically equivalent results independent of physical execution properties, such as the number and size of the workers producing the result. Before the coordinator schedules a planned query pipeline, it checks the result registry for a respective entry. If the entry exists, the coordinator skips the pipeline and repeats this process for the depending query pipelines. If there is no entry, the query pipeline is scheduled for execution, and its result is registered when complete. This cache makes Skyrise more performant and cost-efficient for workloads with recurring queries.

\section{Experimental Evaluation}
\label{section:experimental_evaluation}

In this section, we conduct a series of experiments to answer the following questions about our FaaS-based Skyrise query processor:

\begin{itemize}
    \item How does Skyrise's query performance compare to other cloud-based data analysis systems?
    \item Is Skyrise cost-efficient compared to commercial Query-as-a-Service (QaaS) systems?
    \item How well does Skyrise scale elastically?
\end{itemize}

\subsection{Setup and Methodology}
We conduct all experiments on AWS in the region us-east-1 \cite{aws-zones} in the time frame of August 2024 to January 2025. For our experiments on Skyrise, we deploy ARM-based Lambda functions \cite{lambda-isas} and deactivate the query result cache. Our driver has low resource requirements and runs on an \texttt{c6g.xlarge} instance \cite{ec2-c6g} across experiments. All other compute resources are newly created for each experiment configuration and repetition. As cloud data analysis systems, we consider AWS Athena~\cite{athena}, Snowflake~\cite{snowflake}, and Lambada~\cite{Mueller:2020}. We use queries and datasets from the TPC-H~\cite{tpc-h} benchmark. The tables are partitioned into Parquet files using \texttt{ZSTD} compression and are stored on S3. We employ the standard generators and do not partition or sort on any specific keys. For TPC-H scale factor 1,000, the \texttt{lineitem} table totals 177.4 GiB and the \texttt{orders} table 44.9 GiB. 

\subsection{End-to-End Workloads}
To assess the performance and cost efficiency of Skyrise, we compare it against several state-of-the-art approaches: The academic research prototype Lambada~\cite{Mueller:2020} and the commercial offerings Snowflake and Athena.
\textbf{Lambada} (see \Cref{section:related_work}) is a research prototype for FaaS-based query processing. We report the numbers published in~\cite{Mueller:2020}.
\textbf{Snowflake} is a commercial cloud-based data warehouse offering.
Typically, users allocate several warehouse instances.
Recently, Snowflake added serverless tasks which are mainly aimed at data maintenance.
Serverless tasks are executed on warehouses provisioned by Snowflake and are billed per second.
Serverless tasks are created with an initial warehouse size provided by the user.
After a few executions, Snowflake automatically determines the best warehouse size for each tasks.
For stable measurements, we created new tasks for each query execution and set the warehouse size to \texttt{2xlarge} (the configuration with the best price-performance ratio).
We evaluated five setups for Snowflake (\texttt{SF}): the default of loading data into the proprietary format (\texttt{Server/Intern}), as external tables from warehouses (\texttt{Server/Extern}) and serverless tasks (\texttt{Serverless/Extern}), and as Iceberg tables from warehouses (\texttt{Server/Iceberg}) and serverless tasks (\texttt{Serverless/Iceberg}), respectively.
Both the external tables and the Iceberg tables link the S3-resident Parquets files and execute directly on them.
The main difference between both setups is that Snowflake stores meta data for Iceberg tables.
As Athena, Lambada, and Skyrise do not store any metadata on the table data, we consider the external table setup to be the most relevant for a fair comparison.
We use Snowflake version 8.38.3 and take costs provided by Snowflake's monitoring tools (with a standard configuration costing \$2 per compute credit). For the warehouse-based Snowflake variants, we multiply the credit price with the query runtime.
\textbf{Athena} is a serverless query processing offering by Amazon.
Athena builds on Trino (previously Presto \cite{Sethi:2019}).
Similar to the setup for Snowflake, we created external tables based on the TPC-H dataset.
Resources used for queries are automatically determined by Athena and cannot be adapted or observed by users.
For the determined costs per query, we use the reported bytes scanned by Athena and the current pricing model of \$5 per terabyte of data scanned.

We run the three scan-heavy TPC-H queries 1, 6, and 12 (scale factor 1,000), which are well-suited for serverless execution as they are not dominated by shuffling operators.
We measure runtimes and costs per query execution, reporting the median of five executions.

\subsubsection{Query Latency}
\begin{figure}[bt]
    \centering
    \includegraphics[width=1.0\linewidth]{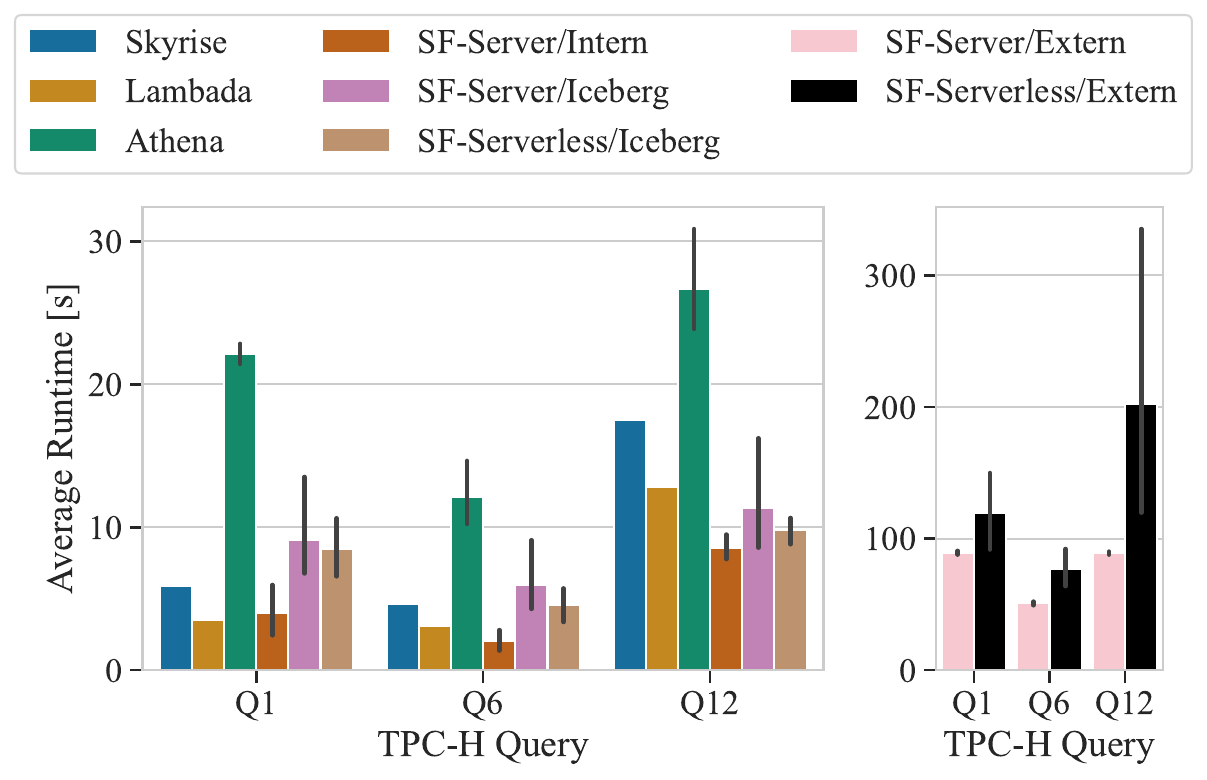}
    \vspace{-0.5cm}
    \caption{Query performance for TPC-H queries.}
    \vspace{-0.3cm}
    \label{figure:system_performance}
\end{figure}

Users with ad-hoc analytical query workloads want interactive response times. We present our results in \figureref{figure:system_performance}.
The overall fastest system is standard Snowflake with its internal data format. It is the only system that does not need to read data from S3.
From the remaining systems, the fastest  for Q1 and Q6 is Lambada, followed by Skyrise, which is on average 30\,\% slower. Both systems share key architecture and design features. They exploit the scalability of their underlying FaaS platform and efficiently handle columnar file formats on remote cloud storage.
For Q12, the two Snowflake variants working with Iceberg tables are faster than the academic prototypes.
They are followed by Athena, which is $2-3 \times$ slower in our evaluation.
Comparing the external table setups of Snowflake, Athena is $5-8 \times$ faster on average. The QaaS systems scale out their resources conservatively in serverless mode, which degrades performance for infrequent terabyte-scale queries. Specifically for Snowflake, the external table setups appear to miss a range of statistics-based optimizations as compared to Iceberg tables.

We hypothesize that the reason for the performance difference between Lambada and Skyrise is the dataset for the Lambada experiments \cite{Mueller:2020}. This dataset differs in two significant ways from ours. First, the string columns are replaced with integers, reducing the time it takes to load and parse them. Second, the \texttt{lineitem} relation is sorted on the \texttt{l\_shipdate} attribute, which enables partition pruning for the evaluated queries.

\subsubsection{Monetary Cost}
\begin{figure}[bt]
    \centering
    \includegraphics[width=1.0\linewidth]{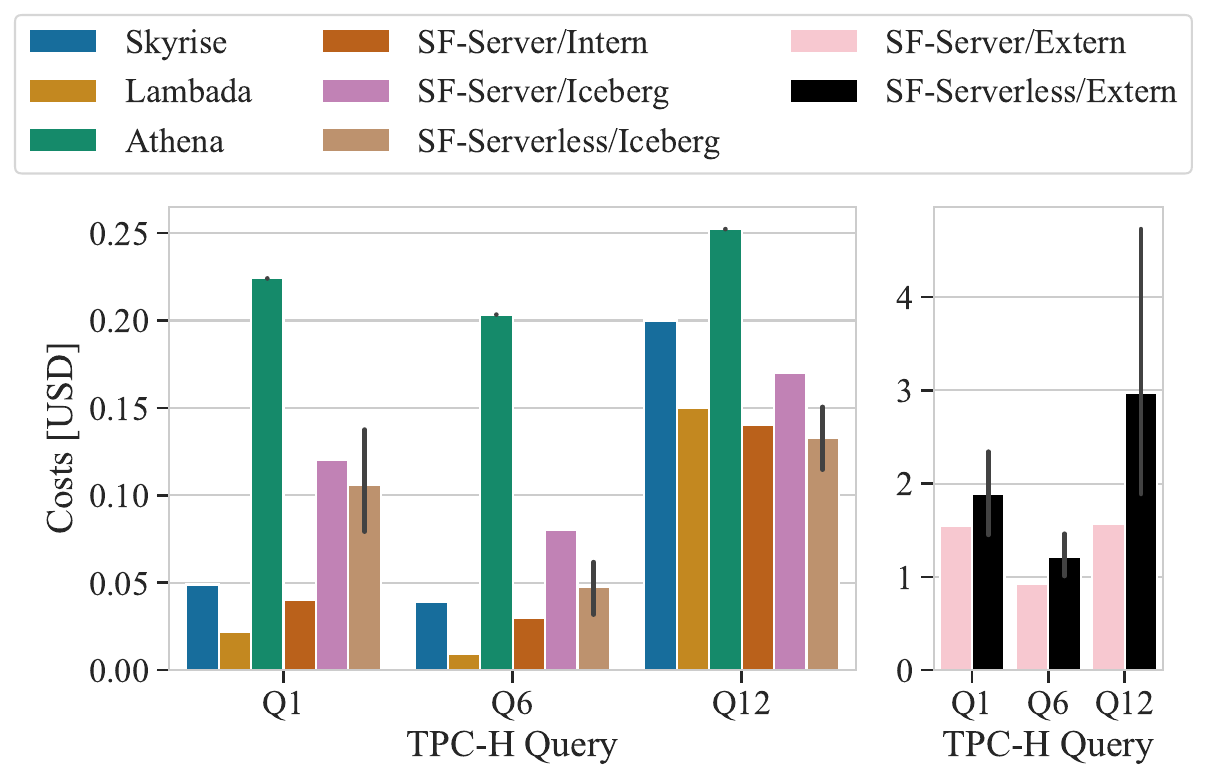}
    \vspace{-0.5cm}
    \caption{Query costs for TPC-H queries.}
    \vspace{-0.3cm}
    \label{figure:system_cost}
\end{figure}

\figureref{figure:system_cost} shows the average costs per query.
The pattern is comparable to the latency results.
Lambada has the lowest costs per query, followed by Skyrise and the Snowflake variants with internal and Iceberg tables.
Athena and Snowflake with external tables follow with Athena being $7 \times$ cheaper on average.

Skyrise achieves comparable latencies and costs as the state-of-the-art system Lambada.
The commercial systems Athena and Snowflake (again with internal and Iceberg tables) closely follow the research systems, which is noteworthy given that the non-commercial systems Lambada and Skyrise only incur infrastructure costs.
In this scenario, the recent Snowflake serverless offering is not cost-competitive with Athena.

\subsection{Elasticity}
\begin{figure}[bt]
    \centering
    \includegraphics[width=1.0\linewidth]{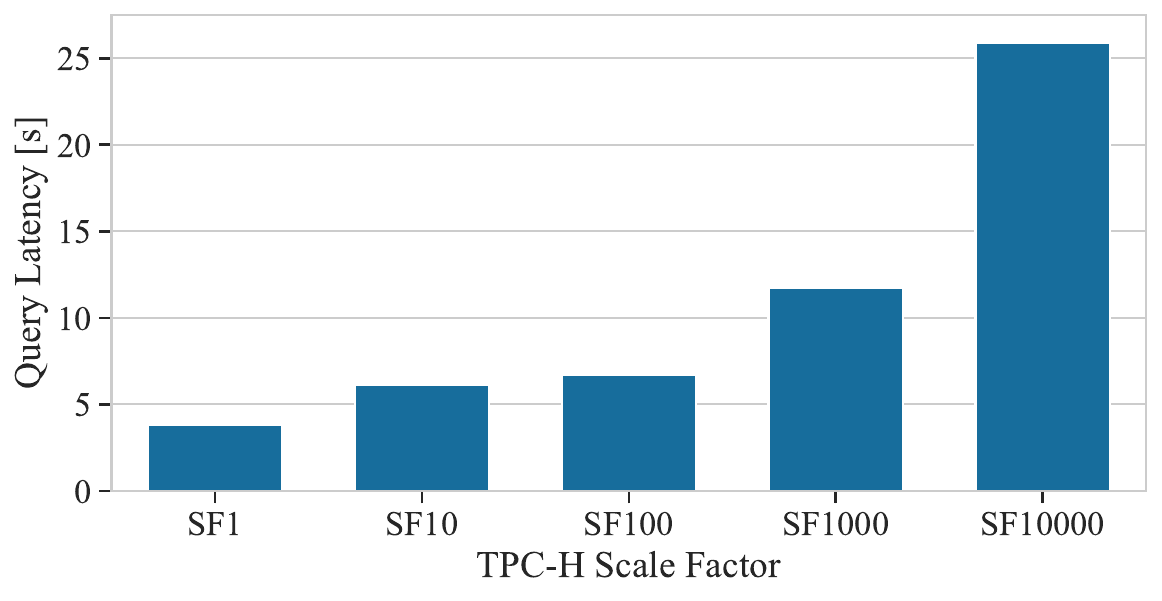}
    \vspace{-0.5cm}
    \caption{Aggregated query latencies for TPC-H queries 1 \& 6.}
    \vspace{-0.3cm}
    \label{figure:skyrise_elasticity}
\end{figure}

To evaluate Skyrise's elasticity, we run the simple scan queries TPC-H Q1 and Q6 at five different scale factors from 1 ({\raise.17ex\hbox{$\scriptstyle\mathtt{\sim}$}}1 GB) to 10,000 ({\raise.17ex\hbox{$\scriptstyle\mathtt{\sim}$}}10 TB). These queries have linear complexity and require resources proportionally to their input data.

We run the queries cold in 15 minute intervals to avoid any caching or provisioning effects in the AWS infrastructure.
The results are shown in \figureref{figure:skyrise_elasticity}. We report the aggregated runtimes of both queries. Given ideal elastic scalability, we would expect the query latency to be the same across all scale factors. We, however, observe that larger scale factors, from 10, but especially 1,000 and 10,000 are generally more prone to stragglers in function invocations and storage requests. The only configuration without straggler effects is scale factor 1, which runs on a single worker. Although Skyrise retriggers stragglers at different levels of the stack, detecting and mitigating them comes with an increasing overhead. This is the case specifically for the 10,000 configuration, which runs with a maximum of 2,500 query workers and suffers from IOPS limits in S3.
Generally, without any upfront provisioning or load, Skyrise automatically scales the resources with the size of the query, and we see less than an order of magnitude (OOM) difference in query latency, despite the problem sizes varying by 5 OOMs. Thus, Skyrise effectively uses the elasticity inherent in its serverless compute and storage resources.

\section{Related Work}
\label{section:related_work}

In recent years, both academia and industry have been interested in serverless data processing, with a particular focus on elasticity. There are research prototypes that build upon FaaS platforms and serverless storage to inherit their elasticity. Other, primarily commercial, systems add techniques to reallocate resources between their tenants in cloud environments. We relate to both lines of research.

\subsection{FaaS-based Data Processing}

Following its initial public release in 2014, FaaS technology was notably adopted by Jonas et al. who developed PyWren \cite{Jonas:2017}. PyWren is a prototypical implementation of a parallel programming model for serverless cloud environments. While PyWren targeted applications more generally, the potential of cloud functions for data analytics was shown by Kim et al. with Flint~\cite{Kim:2018} and later Pu et al. with Locus \cite{Pu:2019}. Flint is a rewrite of Apache Spark's execution layer using serverless building blocks like workers and disaggregated storage. In addition, Locus proposed a shuffling operator using S3 and AWS ElasticCache. Locus was extended by Zhang et al. with a serverless SQL query engine~\cite{Zhang:2021}, called Caerus, focusing on analytical query processing. Skyrise is the first fully serverless system built entirely on serverless infrastructure without the need for servers hosting a driver, coordinator, or shuffle service. Having Skyrise open-source, allows for using it for research and teaching purposes at HPI and beyond.

The shuffling approaches from PyWren, Flint, and Corral \cite{corral} were enhanced by Lambada \cite{Mueller:2020} and Starling \cite{Perron:2020} through staged shuffles, reducing the necessary storage requests for shuffles. Tiered shuffling was designed in Locus, Cackle \cite{Perron:2023}, Pixels~\cite{Bian:2023}, Caerus~/Jiffy~\cite{Khandelwal:2022} and Pocket~\cite{Klimovic:2018}. Boxer~\cite{Wawrzoniak:2021} shuffles data via NAT and Flock~\cite{Liao:2023} for streaming utilizes function invocation payloads~\cite{Liao:2023}. Skyrise is the first system with tiered shuffling to hot serverless S3 Express storage.

Starling \cite{Perron:2020} addresses performance robustness when processing data on S3 through pipelining and solutions for I/O stragglers. Flint \cite{DBLP:conf/eurosys/SharmaGHIS16} employs a fine-granular fault-tolerance scheme with checkpointing and server selection to cope with cheap but transient servers. Salama et al. introduce a cost-based approach to selecting subsets of intermediate results to cache. In contrast, Skyrise tracks worker progress with query context and runtime statistics to prevent straggling workers due to computation or data skew.

Narasayya and Chaudhuri surveyed cloud data services and discussed the challenges for serverless computing~\cite{DBLP:journals/ftdb/NarasayyaC21}.
The authors mention overheads to write worker state due to their short lifetimes or inefficiencies that arise due to missing network connections between workers.
Workloads that require communication (e.g., transaction protocols or shuffles) thus need to write data to slow object stores.
When adapting existing DBMS to serverless setups, the authors list assumptions in current systems that need to be reconsidered and the challenge to balance performance and costs of (almost) infinitely scalable processing capabilities.

Combining stateful and stateless compute is an interesting new area that Pixels \cite{Bian:2023} and Cackle \cite{Perron:2023} introduced as hybrid compute to minize cost. For more cost-efficient serverless processing of high query volumes, Skyrise introduces partial result caching on serverless storage.

\subsection{Elastic Cloud Data Processing}

Presto \cite{Sethi:2019} is an open source query engine for relational and non-relational data.
Trino (previously known as PrestoSQL) is the backend of Amazon's commercial serverless Athena \cite{athena} offering.

Redshift~\cite{Armenatzoglou:2022} is a distributed cloud-based data warehouse by Amazon.
Its companion, Redshift Spectrum allows to run queries serverless directly on object store data.

Dremel is a distributed analytical query processing system developed by Google~\cite{Melnik:2020}.
Dremel is the underlying technology behind BigQuery.

Snowflake~\cite{DBLP:conf/sigmod/DagevilleCZAABC16} is a commercial cloud-based warehouse solution.
Pricing is based on the user-selected warehouse size (per-second billing), storage usage, and egress data.
Recently, Snowflake added capabilities to schedule serverless tasks (see \Cref{section:experimental_evaluation}).

Databricks introduced the concept of \emph{data lakehouses}~\cite{Zaharia:2021}.
Lakehouses are cloud-based analytical systems similar to data warehouses like Snowflake with the main difference that they directly process queries on open direct-access data formats, such as Parquet.
Users can directly query structured, semi-structured, and unstructured data without upfront data ingestion.
Being initially built on Spark, the lakehouse architecture aims to combine machine learning with analytical query processing.

\section{Conclusion} 
\label{section:conclusion}

In this paper, we present the serverless query processor Skyrise. Skyrise is completely built on the serverless premise and fully scalable from zero to thousands of cloud functions, with no stateful coordination. Skyrise is open-source and publicly available. Our experiments show Skyrise's scalability and cost-efficiency for sporadic and variable workloads.

\backmatter

\bmhead{Supplementary information} The source code, data, and/or other artifacts have been made available at \url{https://github.com/hpides/skyrise}.

\bmhead{Acknowledgments}
We thank the Master's students that worked on the Skyrise project and implemented components of the query processor: Tobias Maltenberger, Julian Menzler, Theo Radig, Pascal Schulze, Jan Siebert, and many more.

This work was partially funded by SAP, the AWS Cloud Credit for Research Program, the German Research Foundation (414984028), and the European Union’s Horizon 2020 research and innovation programme (957407).

\bibliography{sn-bibliography}


\begin{thebibliography}{66}
\ifx \bisbn   \undefined \def \bisbn  #1{ISBN #1}\fi
\ifx \binits  \undefined \def \binits#1{#1}\fi
\ifx \bauthor  \undefined \def \bauthor#1{#1}\fi
\ifx \batitle  \undefined \def \batitle#1{#1}\fi
\ifx \bjtitle  \undefined \def \bjtitle#1{#1}\fi
\ifx \bvolume  \undefined \def \bvolume#1{\textbf{#1}}\fi
\ifx \byear  \undefined \def \byear#1{#1}\fi
\ifx \bissue  \undefined \def \bissue#1{#1}\fi
\ifx \bfpage  \undefined \def \bfpage#1{#1}\fi
\ifx \blpage  \undefined \def \blpage #1{#1}\fi
\ifx \burl  \undefined \def \burl#1{\textsf{#1}}\fi
\ifx \doiurl  \undefined \def \doiurl#1{\url{https://doi.org/#1}}\fi
\ifx \betal  \undefined \def \betal{\textit{et al.}}\fi
\ifx \binstitute  \undefined \def \binstitute#1{#1}\fi
\ifx \binstitutionaled  \undefined \def \binstitutionaled#1{#1}\fi
\ifx \bctitle  \undefined \def \bctitle#1{#1}\fi
\ifx \beditor  \undefined \def \beditor#1{#1}\fi
\ifx \bpublisher  \undefined \def \bpublisher#1{#1}\fi
\ifx \bbtitle  \undefined \def \bbtitle#1{#1}\fi
\ifx \bedition  \undefined \def \bedition#1{#1}\fi
\ifx \bseriesno  \undefined \def \bseriesno#1{#1}\fi
\ifx \blocation  \undefined \def \blocation#1{#1}\fi
\ifx \bsertitle  \undefined \def \bsertitle#1{#1}\fi
\ifx \bsnm \undefined \def \bsnm#1{#1}\fi
\ifx \bsuffix \undefined \def \bsuffix#1{#1}\fi
\ifx \bparticle \undefined \def \bparticle#1{#1}\fi
\ifx \barticle \undefined \def \barticle#1{#1}\fi
\bibcommenthead
\ifx \bconfdate \undefined \def \bconfdate #1{#1}\fi
\ifx \botherref \undefined \def \botherref #1{#1}\fi
\ifx \url \undefined \def \url#1{\textsf{#1}}\fi
\ifx \bchapter \undefined \def \bchapter#1{#1}\fi
\ifx \bbook \undefined \def \bbook#1{#1}\fi
\ifx \bcomment \undefined \def \bcomment#1{#1}\fi
\ifx \oauthor \undefined \def \oauthor#1{#1}\fi
\ifx \citeauthoryear \undefined \def \citeauthoryear#1{#1}\fi
\ifx \endbibitem  \undefined \def \endbibitem {}\fi
\ifx \bconflocation  \undefined \def \bconflocation#1{#1}\fi
\ifx \arxivurl  \undefined \def \arxivurl#1{\textsf{#1}}\fi
\csname PreBibitemsHook\endcsname

\bibitem[\protect\citeauthoryear{Gupta et~al.}{2015}]{Gupta:2015}
\begin{bchapter}
\bauthor{\bsnm{Gupta}, \binits{A.}},
\bauthor{\bsnm{Agarwal}, \binits{D.}},
\bauthor{\bsnm{Tan}, \binits{D.}},
\bauthor{\bsnm{Kulesza}, \binits{J.}},
\bauthor{\bsnm{Pathak}, \binits{R.}},
\bauthor{\bsnm{Stefani}, \binits{S.}},
\bauthor{\bsnm{Srinivasan}, \binits{V.}}:
\bctitle{{Amazon Redshift and the Case for Simpler Data Warehouses}}.
In: \bbtitle{{ACM} {SIGMOD}},
pp. \bfpage{1917}--\blpage{1923}
(\byear{2015})
\end{bchapter}
\endbibitem

\bibitem[\protect\citeauthoryear{Melnik et~al.}{2010}]{Melnik:2010}
\begin{barticle}
\bauthor{\bsnm{Melnik}, \binits{S.}},
\bauthor{\bsnm{Gubarev}, \binits{A.}},
\bauthor{\bsnm{Long}, \binits{J.J.}},
\bauthor{\bsnm{Romer}, \binits{G.}},
\bauthor{\bsnm{Shivakumar}, \binits{S.}},
\bauthor{\bsnm{Tolton}, \binits{M.}},
\bauthor{\bsnm{Vassilakis}, \binits{T.}}:
\batitle{{Dremel: Interactive Analysis of Web-Scale Datasets}}.
\bjtitle{{PVLDB}}
\bvolume{3}(\bissue{1}),
\bfpage{330}--\blpage{339}
(\byear{2010})
\end{barticle}
\endbibitem

\bibitem[\protect\citeauthoryear{Dageville et~al.}{2016}]{Dageville:2016}
\begin{bchapter}
\bauthor{\bsnm{Dageville}, \binits{B.}},
\bauthor{\bsnm{Cruanes}, \binits{T.}},
\bauthor{\bsnm{Zukowski}, \binits{M.}},
\bauthor{\bsnm{Antonov}, \binits{V.}},
\bauthor{\bsnm{Avanes}, \binits{A.}},
\bauthor{\bsnm{Bock}, \binits{J.}},
\bauthor{\bsnm{Claybaugh}, \binits{J.}},
\bauthor{\bsnm{Engovatov}, \binits{D.}},
\bauthor{\bsnm{Hentschel}, \binits{M.}},
\bauthor{\bsnm{Huang}, \binits{J.}},
\bauthor{\bsnm{Lee}, \binits{A.W.}},
\bauthor{\bsnm{Motivala}, \binits{A.}},
\bauthor{\bsnm{Munir}, \binits{A.Q.}},
\bauthor{\bsnm{Pelley}, \binits{S.}},
\bauthor{\bsnm{Povinec}, \binits{P.}},
\bauthor{\bsnm{Rahn}, \binits{G.}},
\bauthor{\bsnm{Triantafyllis}, \binits{S.}},
\bauthor{\bsnm{Unterbrunner}, \binits{P.}}:
\bctitle{{The Snowflake Elastic Data Warehouse}}.
In: \bbtitle{{ACM} {SIGMOD}},
pp. \bfpage{215}--\blpage{226}
(\byear{2016})
\end{bchapter}
\endbibitem

\bibitem[\protect\citeauthoryear{{IT Market Strategy}}{2022}]{dbms-market-2021}
\begin{botherref}
\oauthor{\bsnm{{IT Market Strategy}}}:
{DBMS Market Transformation 2021: The Big Picture}.
\url{https://itmarketstrategy.com/2022/12/09/dbms-market-transformation-2021-the-big-picture}.
Accessed: 2024-10-07
(2022)
\end{botherref}
\endbibitem

\bibitem[\protect\citeauthoryear{{IT Market Strategy}}{2024}]{dbms-market-2023}
\begin{botherref}
\oauthor{\bsnm{{IT Market Strategy}}}:
{DBMS Market 2023 – More Momentum Shifts}.
\url{https://itmarketstrategy.com/2024/05/31/dbms-market-2023-more-momentum-shifts}.
Accessed: 2024-10-07
(2024)
\end{botherref}
\endbibitem

\bibitem[\protect\citeauthoryear{Singh et~al.}{2015}]{Singh:2015}
\begin{bchapter}
\bauthor{\bsnm{Singh}, \binits{A.}},
\bauthor{\bsnm{Ong}, \binits{J.}},
\bauthor{\bsnm{Agarwal}, \binits{A.}},
\bauthor{\bsnm{Anderson}, \binits{G.}},
\bauthor{\bsnm{Armistead}, \binits{A.}},
\bauthor{\bsnm{Bannon}, \binits{R.}},
\bauthor{\bsnm{Boving}, \binits{S.}},
\bauthor{\bsnm{Desai}, \binits{G.}},
\bauthor{\bsnm{Felderman}, \binits{B.}},
\bauthor{\bsnm{Germano}, \binits{P.}},
\bauthor{\bsnm{Kanagala}, \binits{A.}},
\bauthor{\bsnm{Provost}, \binits{J.}},
\bauthor{\bsnm{Simmons}, \binits{J.}},
\bauthor{\bsnm{Tanda}, \binits{E.}},
\bauthor{\bsnm{Wanderer}, \binits{J.}},
\bauthor{\bsnm{H{\"{o}}lzle}, \binits{U.}},
\bauthor{\bsnm{Stuart}, \binits{S.}},
\bauthor{\bsnm{Vahdat}, \binits{A.}}:
\bctitle{{Jupiter Rising: {A} Decade of Clos Topologies and Centralized Control in Google's Datacenter Network}}.
In: \bbtitle{{ACM} {SIGCOMM}},
pp. \bfpage{183}--\blpage{197}
(\byear{2015})
\end{bchapter}
\endbibitem

\bibitem[\protect\citeauthoryear{Firestone et~al.}{2018}]{Firestone:2018}
\begin{bchapter}
\bauthor{\bsnm{Firestone}, \binits{D.}},
\bauthor{\bsnm{Putnam}, \binits{A.}},
\bauthor{\bsnm{Mundkur}, \binits{S.}},
\bauthor{\bsnm{Chiou}, \binits{D.}},
\bauthor{\bsnm{Dabagh}, \binits{A.}},
\bauthor{\bsnm{Andrewartha}, \binits{M.}},
\bauthor{\bsnm{Angepat}, \binits{H.}},
\bauthor{\bsnm{Bhanu}, \binits{V.}},
\bauthor{\bsnm{Caulfield}, \binits{A.M.}},
\bauthor{\bsnm{Chung}, \binits{E.S.}},
\bauthor{\bsnm{Chandrappa}, \binits{H.K.}},
\bauthor{\bsnm{Chaturmohta}, \binits{S.}},
\bauthor{\bsnm{Humphrey}, \binits{M.}},
\bauthor{\bsnm{Lavier}, \binits{J.}},
\bauthor{\bsnm{Lam}, \binits{N.}},
\bauthor{\bsnm{Liu}, \binits{F.}},
\bauthor{\bsnm{Ovtcharov}, \binits{K.}},
\bauthor{\bsnm{Padhye}, \binits{J.}},
\bauthor{\bsnm{Popuri}, \binits{G.}},
\bauthor{\bsnm{Raindel}, \binits{S.}},
\bauthor{\bsnm{Sapre}, \binits{T.}},
\bauthor{\bsnm{Shaw}, \binits{M.}},
\bauthor{\bsnm{Silva}, \binits{G.}},
\bauthor{\bsnm{Sivakumar}, \binits{M.}},
\bauthor{\bsnm{Srivastava}, \binits{N.}},
\bauthor{\bsnm{Verma}, \binits{A.}},
\bauthor{\bsnm{Zuhair}, \binits{Q.}},
\bauthor{\bsnm{Bansal}, \binits{D.}},
\bauthor{\bsnm{Burger}, \binits{D.}},
\bauthor{\bsnm{Vaid}, \binits{K.}},
\bauthor{\bsnm{Maltz}, \binits{D.A.}},
\bauthor{\bsnm{Greenberg}, \binits{A.G.}}:
\bctitle{{Azure Accelerated Networking: SmartNICs in the Public Cloud}}.
In: \bbtitle{{USENIX} {NSDI}},
pp. \bfpage{51}--\blpage{66}
(\byear{2018})
\end{bchapter}
\endbibitem

\bibitem[\protect\citeauthoryear{{Amazon}}{2024}]{nitro}
\begin{botherref}
\oauthor{\bsnm{{Amazon}}}:
{AWS Nitro System}.
\url{https://aws.amazon.com/ec2/nitro/}.
Accessed: 2024-10-07
(2024)
\end{botherref}
\endbibitem

\bibitem[\protect\citeauthoryear{{Amazon}}{2017}]{ec2-pricing-fine}
\begin{botherref}
\oauthor{\bsnm{{Amazon}}}:
{Per-Second Billing for EC2 Instances and EBS Volumes}.
\url{https://aws.amazon.com/blogs/aws/new-per-second-billing-for-ec2-instances-and-ebs-volumes}.
Accessed: 2024-10-07
(2017)
\end{botherref}
\endbibitem

\bibitem[\protect\citeauthoryear{{Google}}{2024}]{gcp-pricing}
\begin{botherref}
\oauthor{\bsnm{{Google}}}:
{Google Cloud Pricing}.
\url{https://cloud.google.com/pricing/}.
Accessed: 2024-10-07
(2024)
\end{botherref}
\endbibitem

\bibitem[\protect\citeauthoryear{Armenatzoglou et~al.}{2022}]{Armenatzoglou:2022}
\begin{bchapter}
\bauthor{\bsnm{Armenatzoglou}, \binits{N.}},
\bauthor{\bsnm{Basu}, \binits{S.}},
\bauthor{\bsnm{Bhanoori}, \binits{N.}},
\bauthor{\bsnm{Cai}, \binits{M.}},
\bauthor{\bsnm{Chainani}, \binits{N.}},
\bauthor{\bsnm{Chinta}, \binits{K.}},
\bauthor{\bsnm{Govindaraju}, \binits{V.}},
\bauthor{\bsnm{Green}, \binits{T.J.}},
\bauthor{\bsnm{Gupta}, \binits{M.}},
\bauthor{\bsnm{Hillig}, \binits{S.}},
\bauthor{\bsnm{Hotinger}, \binits{E.}},
\bauthor{\bsnm{Leshinksy}, \binits{Y.}},
\bauthor{\bsnm{Liang}, \binits{J.}},
\bauthor{\bsnm{McCreedy}, \binits{M.}},
\bauthor{\bsnm{Nagel}, \binits{F.}},
\bauthor{\bsnm{Pandis}, \binits{I.}},
\bauthor{\bsnm{Parchas}, \binits{P.}},
\bauthor{\bsnm{Pathak}, \binits{R.}},
\bauthor{\bsnm{Polychroniou}, \binits{O.}},
\bauthor{\bsnm{Rahman}, \binits{F.}},
\bauthor{\bsnm{Saxena}, \binits{G.}},
\bauthor{\bsnm{Soundararajan}, \binits{G.}},
\bauthor{\bsnm{Subramanian}, \binits{S.}},
\bauthor{\bsnm{Terry}, \binits{D.}}:
\bctitle{{Amazon Redshift Re-invented}}.
In: \bbtitle{{ACM} {SIGMOD}},
pp. \bfpage{2205}--\blpage{2217}
(\byear{2022})
\end{bchapter}
\endbibitem

\bibitem[\protect\citeauthoryear{Melnik et~al.}{2020}]{Melnik:2020}
\begin{barticle}
\bauthor{\bsnm{Melnik}, \binits{S.}},
\bauthor{\bsnm{Gubarev}, \binits{A.}},
\bauthor{\bsnm{Long}, \binits{J.J.}},
\bauthor{\bsnm{Romer}, \binits{G.}},
\bauthor{\bsnm{Shivakumar}, \binits{S.}},
\bauthor{\bsnm{Tolton}, \binits{M.}},
\bauthor{\bsnm{Vassilakis}, \binits{T.}},
\bauthor{\bsnm{Ahmadi}, \binits{H.}},
\bauthor{\bsnm{Delorey}, \binits{D.}},
\bauthor{\bsnm{Min}, \binits{S.}},
\bauthor{\bsnm{Pasumansky}, \binits{M.}},
\bauthor{\bsnm{Shute}, \binits{J.}}:
\batitle{{Dremel: {A} Decade of Interactive {SQL} Analysis at Web Scale}}.
\bjtitle{{PVLDB}}
\bvolume{13}(\bissue{12}),
\bfpage{3461}--\blpage{3472}
(\byear{2020})
\end{barticle}
\endbibitem

\bibitem[\protect\citeauthoryear{Vuppalapati et~al.}{2020}]{Vuppalapati:2020}
\begin{bchapter}
\bauthor{\bsnm{Vuppalapati}, \binits{M.}},
\bauthor{\bsnm{Miron}, \binits{J.}},
\bauthor{\bsnm{Agarwal}, \binits{R.}},
\bauthor{\bsnm{Truong}, \binits{D.}},
\bauthor{\bsnm{Motivala}, \binits{A.}},
\bauthor{\bsnm{Cruanes}, \binits{T.}}:
\bctitle{{Building An Elastic Query Engine on Disaggregated Storage}}.
In: \bbtitle{{USENIX} {NSDI}},
pp. \bfpage{449}--\blpage{462}
(\byear{2020})
\end{bchapter}
\endbibitem

\bibitem[\protect\citeauthoryear{{Amazon}}{2023}]{lambda}
\begin{botherref}
\oauthor{\bsnm{{Amazon}}}:
{AWS Lambda}.
\url{https://aws.amazon.com/lambda/}.
Accessed: 2024-10-07
(2023)
\end{botherref}
\endbibitem

\bibitem[\protect\citeauthoryear{{Microsoft}}{2024}]{maf}
\begin{botherref}
\oauthor{\bsnm{{Microsoft}}}:
{Azure Functions: Execute Event-driven Serverless Code with an End-to-end Development Experience.}
\url{https://azure.microsoft.com/services/functions/}.
Accessed: 2024-03-27
(2024)
\end{botherref}
\endbibitem

\bibitem[\protect\citeauthoryear{Jonas et~al.}{2017}]{Jonas:2017}
\begin{bchapter}
\bauthor{\bsnm{Jonas}, \binits{E.}},
\bauthor{\bsnm{Pu}, \binits{Q.}},
\bauthor{\bsnm{Venkataraman}, \binits{S.}},
\bauthor{\bsnm{Stoica}, \binits{I.}},
\bauthor{\bsnm{Recht}, \binits{B.}}:
\bctitle{{Occupy the Cloud: Distributed Computing for the 99\%}}.
In: \bbtitle{{ACM} {SoCC}},
pp. \bfpage{445}--\blpage{451}
(\byear{2017})
\end{bchapter}
\endbibitem

\bibitem[\protect\citeauthoryear{Ustiugov et~al.}{2021}]{Ustiugov:2021}
\begin{bchapter}
\bauthor{\bsnm{Ustiugov}, \binits{D.}},
\bauthor{\bsnm{Amariucai}, \binits{T.}},
\bauthor{\bsnm{Grot}, \binits{B.}}:
\bctitle{{Analyzing Tail Latency in Serverless Clouds with STeLLAR}}.
In: \bbtitle{{IEEE} {IISWC}},
pp. \bfpage{51}--\blpage{62}
(\byear{2021})
\end{bchapter}
\endbibitem

\bibitem[\protect\citeauthoryear{Seemakhupt et~al.}{2023}]{Seemakhupt:2023}
\begin{bchapter}
\bauthor{\bsnm{Seemakhupt}, \binits{K.}},
\bauthor{\bsnm{Stephens}, \binits{B.E.}},
\bauthor{\bsnm{Khan}, \binits{S.M.}},
\bauthor{\bsnm{Liu}, \binits{S.}},
\bauthor{\bsnm{Wassel}, \binits{H.M.G.}},
\bauthor{\bsnm{Yeganeh}, \binits{S.H.}},
\bauthor{\bsnm{Snoeren}, \binits{A.C.}},
\bauthor{\bsnm{Krishnamurthy}, \binits{A.}},
\bauthor{\bsnm{Culler}, \binits{D.E.}},
\bauthor{\bsnm{Levy}, \binits{H.M.}}:
\bctitle{{A Cloud-Scale Characterization of Remote Procedure Calls}}.
In: \bbtitle{{ACM} {SOSP}},
pp. \bfpage{498}--\blpage{514}
(\byear{2023})
\end{bchapter}
\endbibitem

\bibitem[\protect\citeauthoryear{{Parquet contributors}}{2024}]{parquet}
\begin{botherref}
\oauthor{\bsnm{{Parquet contributors}}}:
{Apache Parquet}.
\url{https://parquet.apache.org/}.
Accessed: 2024-10-07
(2024)
\end{botherref}
\endbibitem

\bibitem[\protect\citeauthoryear{{Amazon}}{2024}]{aws-serverless}
\begin{botherref}
\oauthor{\bsnm{{Amazon}}}:
{Serverless on AWS}.
\url{https://aws.amazon.com/serverless/}.
Accessed: 2024-03-27
(2024)
\end{botherref}
\endbibitem

\bibitem[\protect\citeauthoryear{{Google}}{2024}]{gcp-serverless}
\begin{botherref}
\oauthor{\bsnm{{Google}}}:
{Serverless}.
\url{https://cloud.google.com/serverless}.
Accessed: 2024-03-27
(2024)
\end{botherref}
\endbibitem

\bibitem[\protect\citeauthoryear{M{\"{u}}ller et~al.}{2020}]{Mueller:2020}
\begin{bchapter}
\bauthor{\bsnm{M{\"{u}}ller}, \binits{I.}},
\bauthor{\bsnm{Marroqu{\'{\i}}n}, \binits{R.}},
\bauthor{\bsnm{Alonso}, \binits{G.}}:
\bctitle{{Lambada: Interactive Data Analytics on Cold Data Using Serverless Cloud Infrastructure}}.
In: \bbtitle{{ACM} {SIGMOD}},
pp. \bfpage{115}--\blpage{130}
(\byear{2020})
\end{bchapter}
\endbibitem

\bibitem[\protect\citeauthoryear{Perron et~al.}{2020}]{Perron:2020}
\begin{bchapter}
\bauthor{\bsnm{Perron}, \binits{M.}},
\bauthor{\bsnm{Fernandez}, \binits{R.C.}},
\bauthor{\bsnm{DeWitt}, \binits{D.J.}},
\bauthor{\bsnm{Madden}, \binits{S.}}:
\bctitle{{Starling: A Scalable Query Engine on Cloud Functions}}.
In: \bbtitle{{ACM} {SIGMOD}},
pp. \bfpage{131}--\blpage{141}
(\byear{2020})
\end{bchapter}
\endbibitem

\bibitem[\protect\citeauthoryear{{Amazon}}{2024a}]{athena}
\begin{botherref}
\oauthor{\bsnm{{Amazon}}}:
{Amazon Athena: Analyze Petabyte-scale Data Where it Lives with Ease and Flexibility}.
\url{https://aws.amazon.com/athena/}.
Accessed: 2024-10-07
(2024)
\end{botherref}
\endbibitem

\bibitem[\protect\citeauthoryear{{Amazon}}{2024b}]{redshift-serverless}
\begin{botherref}
\oauthor{\bsnm{{Amazon}}}:
{Amazon Redshift Serverless}.
\url{https://aws.amazon.com/redshift/redshift-serverless/}.
Accessed: 2024-10-07
(2024)
\end{botherref}
\endbibitem

\bibitem[\protect\citeauthoryear{{Snowflake}}{2024}]{snowflake-serverless}
\begin{botherref}
\oauthor{\bsnm{{Snowflake}}}:
{Serverless Tasks}.
\url{https://docs.snowflake.com/en/user-guide/tasks-intro#serverless-tasks}.
Accessed: 2024-10-07
(2024)
\end{botherref}
\endbibitem

\bibitem[\protect\citeauthoryear{{Amazon}}{2023}]{s3}
\begin{botherref}
\oauthor{\bsnm{{Amazon}}}:
{Amazon S3}.
\url{https://aws.amazon.com/s3/}.
Accessed: 2024-10-07
(2023)
\end{botherref}
\endbibitem

\bibitem[\protect\citeauthoryear{{Google}}{2024}]{gcp-storage}
\begin{botherref}
\oauthor{\bsnm{{Google}}}:
{Cloud Storage}.
\url{https://cloud.google.com/storage/}.
Accessed: 2024-03-27
(2024)
\end{botherref}
\endbibitem

\bibitem[\protect\citeauthoryear{{Amazon}}{2024}]{lambda-limits}
\begin{botherref}
\oauthor{\bsnm{{Amazon}}}:
{Lambda Quotas}.
\url{https://docs.aws.amazon.com/lambda/latest/dg/gettingstarted-limits.html}.
Accessed: 2024-03-28
(2024)
\end{botherref}
\endbibitem

\bibitem[\protect\citeauthoryear{{Amazon}}{2023}]{ec2-c6g}
\begin{botherref}
\oauthor{\bsnm{{Amazon}}}:
{Amazon EC2 C6g Instances}.
\url{https://aws.amazon.com/ec2/instance-types/c6g/}.
Accessed: 2024-03-27
(2023)
\end{botherref}
\endbibitem

\bibitem[\protect\citeauthoryear{{Microsoft}}{2024}]{mabs}
\begin{botherref}
\oauthor{\bsnm{{Microsoft}}}:
{Azure Blob Storage: Massively Scalable and Secure Object Storage for Cloud-native Workloads, Archives, Data Lakes, High-performance Computing, and Machine Learning.}
\url{https://azure.microsoft.com/products/storage/blobs/}.
Accessed: 2024-03-27
(2024)
\end{botherref}
\endbibitem

\bibitem[\protect\citeauthoryear{{Amazon}}{2018}]{ddb-serverless}
\begin{botherref}
\oauthor{\bsnm{{Amazon}}}:
{Announcing Amazon DynamoDB On-Demand}.
\url{https://aws.amazon.com/about-aws/whats-new/2018/11/announcing-amazon-dynamodb-on-demand/}.
Accessed: 2024-03-26
(2018)
\end{botherref}
\endbibitem

\bibitem[\protect\citeauthoryear{{Google}}{2024}]{gcp-firestore}
\begin{botherref}
\oauthor{\bsnm{{Google}}}:
{Cloud Firestore}.
\url{https://cloud.google.com/firestore/}.
Accessed: 2024-03-27
(2024)
\end{botherref}
\endbibitem

\bibitem[\protect\citeauthoryear{{Amazon}}{2022}]{efs-serverless}
\begin{botherref}
\oauthor{\bsnm{{Amazon}}}:
{Announcing Amazon EFS Elastic Throughput}.
\url{https://aws.amazon.com/blogs/aws/new-announcing-amazon-efs-elastic-throughput/}.
Accessed: 2024-03-26
(2022)
\end{botherref}
\endbibitem

\bibitem[\protect\citeauthoryear{{Microsoft}}{2024}]{mafiles}
\begin{botherref}
\oauthor{\bsnm{{Microsoft}}}:
{Azure Files: Simple, Secure, and Serverless Enterprise-grade Cloud File Shares}.
\url{https://azure.microsoft.com/products/storage/files/}.
Accessed: 2024-03-27
(2024)
\end{botherref}
\endbibitem

\bibitem[\protect\citeauthoryear{Bodner}{2020}]{Bodner:2020}
\begin{bchapter}
\bauthor{\bsnm{Bodner}, \binits{T.}}:
\bctitle{{Elastic Query Processing on Function as a Service Platforms}}.
In: \bbtitle{VLDB PhD Workshop}
(\byear{2020})
\end{bchapter}
\endbibitem

\bibitem[\protect\citeauthoryear{Dreseler et~al.}{2019}]{Dreseler:2019}
\begin{bchapter}
\bauthor{\bsnm{Dreseler}, \binits{M.}},
\bauthor{\bsnm{Kossmann}, \binits{J.}},
\bauthor{\bsnm{Boissier}, \binits{M.}},
\bauthor{\bsnm{Klauck}, \binits{S.}},
\bauthor{\bsnm{Uflacker}, \binits{M.}},
\bauthor{\bsnm{Plattner}, \binits{H.}}:
\bctitle{{Hyrise Re-engineered: An Extensible Database System for Research in Relational In-Memory Data Management}}.
In: \bbtitle{{EDBT}},
pp. \bfpage{313}--\blpage{324}
(\byear{2019})
\end{bchapter}
\endbibitem

\bibitem[\protect\citeauthoryear{{Datadog}}{2023}]{datadog-report}
\begin{botherref}
\oauthor{\bsnm{{Datadog}}}:
{The State of Serverless 2023}.
\url{https://www.datadoghq.com/state-of-serverless/}.
Accessed: 2024-03-28
(2023)
\end{botherref}
\endbibitem

\bibitem[\protect\citeauthoryear{Scheuner and Leitner}{2020}]{Scheuner:2020}
\begin{barticle}
\bauthor{\bsnm{Scheuner}, \binits{J.}},
\bauthor{\bsnm{Leitner}, \binits{P.}}:
\batitle{{Function-as-a-Service Performance Evaluation - A Multivocal Literature Review}}.
\bjtitle{Journal of Systems and Software}
\bvolume{170},
\bfpage{110708}
(\byear{2020})
\end{barticle}
\endbibitem

\bibitem[\protect\citeauthoryear{{Amazon}}{2024}]{lambda-urls}
\begin{botherref}
\oauthor{\bsnm{{Amazon}}}:
{Invoking Lambda Function URLs}.
\url{https://docs.aws.amazon.com/lambda/latest/dg/urls-invocation.html}.
Accessed: 2024-03-25
(2024)
\end{botherref}
\endbibitem

\bibitem[\protect\citeauthoryear{Dreseler et~al.}{2020}]{Dreseler:2020}
\begin{barticle}
\bauthor{\bsnm{Dreseler}, \binits{M.}},
\bauthor{\bsnm{Boissier}, \binits{M.}},
\bauthor{\bsnm{Rabl}, \binits{T.}},
\bauthor{\bsnm{Uflacker}, \binits{M.}}:
\batitle{{Quantifying TPC-H Choke Points and Their Optimizations}}.
\bjtitle{PVLDB}
\bvolume{13}(\bissue{8}),
\bfpage{1206}--\blpage{1220}
(\byear{2020})
\end{barticle}
\endbibitem

\bibitem[\protect\citeauthoryear{Bodner et~al.}{2025}]{Bodner:2025a}
\begin{botherref}
\oauthor{\bsnm{Bodner}, \binits{T.}},
\oauthor{\bsnm{Radig}, \binits{T.}},
\oauthor{\bsnm{Justen}, \binits{D.}},
\oauthor{\bsnm{Ritter}, \binits{D.}},
\oauthor{\bsnm{Rabl}, \binits{T.}}:
{An Empirical Evaluation of Serverless Cloud Infrastructure for Large-Scale Data Processing}.
arXiv preprint arXiv:2501.07771
(2025)
{\href{https://arxiv.org/abs/2501.07771}{{arXiv:2501.07771}}}
{[cs.DB]}
\end{botherref}
\endbibitem

\bibitem[\protect\citeauthoryear{{Amazon}}{2024}]{s3-eoz}
\begin{botherref}
\oauthor{\bsnm{{Amazon}}}:
{Amazon S3 Express One Zone Storage Class}.
\url{https://aws.amazon.com/s3/storage-classes/express-one-zone/}.
Accessed: 2024-03-25
(2024)
\end{botherref}
\endbibitem

\bibitem[\protect\citeauthoryear{Mahling et~al.}{2023}]{Mahling:2023}
\begin{bchapter}
\bauthor{\bsnm{Mahling}, \binits{F.}},
\bauthor{\bsnm{R{\"{o}}{\ss}ler}, \binits{P.}},
\bauthor{\bsnm{Bodner}, \binits{T.}},
\bauthor{\bsnm{Rabl}, \binits{T.}}:
\bctitle{{BabelMR: {A} Polyglot Framework for Serverless MapReduce}}.
In: \bbtitle{Joint Proceedings of Workshops at the 49th International Conference on Very Large Data Bases}.
\bsertitle{{CEUR} Workshop Proceedings},
vol. \bseriesno{3462}
(\byear{2023}).
\burl{https://ceur-ws.org/Vol-3462/SDA2.pdf}
\end{bchapter}
\endbibitem

\bibitem[\protect\citeauthoryear{Durner et~al.}{2023}]{Durner:2023}
\begin{barticle}
\bauthor{\bsnm{Durner}, \binits{D.}},
\bauthor{\bsnm{Leis}, \binits{V.}},
\bauthor{\bsnm{Neumann}, \binits{T.}}:
\batitle{{Exploiting Cloud Object Storage for High-Performance Analytics}}.
\bjtitle{PVLDB}
\bvolume{16}(\bissue{11}),
\bfpage{2769}--\blpage{2782}
(\byear{2023})
\end{barticle}
\endbibitem

\bibitem[\protect\citeauthoryear{Ailamaki et~al.}{2002}]{Ailamaki:2002}
\begin{barticle}
\bauthor{\bsnm{Ailamaki}, \binits{A.}},
\bauthor{\bsnm{DeWitt}, \binits{D.J.}},
\bauthor{\bsnm{Hill}, \binits{M.D.}}:
\batitle{{Data Page Layouts for Relational Databases on Deep Memory Hierarchies}}.
\bjtitle{{VLDB} J.}
\bvolume{11}(\bissue{3}),
\bfpage{198}--\blpage{215}
(\byear{2002})
\end{barticle}
\endbibitem

\bibitem[\protect\citeauthoryear{{ORC contributors}}{2024}]{orc}
\begin{botherref}
\oauthor{\bsnm{{ORC contributors}}}:
{Apache ORC}.
\url{https://orc.apache.org/}.
Accessed: 2024-03-26
(2024)
\end{botherref}
\endbibitem

\bibitem[\protect\citeauthoryear{{Amazon}}{2024a}]{aws-zones}
\begin{botherref}
\oauthor{\bsnm{{Amazon}}}:
{Regions and Zones}.
\url{https://docs.aws.amazon.com/AWSEC2/latest/UserGuide/using-regions-availability-zones.html\#concepts-availability-zones}.
Accessed: 2024-03-28
(2024)
\end{botherref}
\endbibitem

\bibitem[\protect\citeauthoryear{{Amazon}}{2024b}]{lambda-isas}
\begin{botherref}
\oauthor{\bsnm{{Amazon}}}:
{Lambda Instruction Set Architectures (ARM/x86)}.
\url{https://docs.aws.amazon.com/lambda/latest/dg/foundation-arch.html}.
Accessed: 2024-03-27
(2024)
\end{botherref}
\endbibitem

\bibitem[\protect\citeauthoryear{{Snowflake}}{2024}]{snowflake}
\begin{botherref}
\oauthor{\bsnm{{Snowflake}}}:
{Snowflake Cloud Data Platform}.
\url{https://www.snowflake.com/de/data-cloud/platform/}.
Accessed: 2024-10-07
(2024)
\end{botherref}
\endbibitem

\bibitem[\protect\citeauthoryear{{Transaction Processing Performance Council}}{2024}]{tpc-h}
\begin{botherref}
\oauthor{\bsnm{{Transaction Processing Performance Council}}}:
{Specification of the TPC-H Benchmark}.
\url{https://www.tpc.org/tpch/}.
Accessed: 2024-03-27
(2024)
\end{botherref}
\endbibitem

\bibitem[\protect\citeauthoryear{Sethi et~al.}{2019}]{Sethi:2019}
\begin{bchapter}
\bauthor{\bsnm{Sethi}, \binits{R.}},
\bauthor{\bsnm{Traverso}, \binits{M.}},
\bauthor{\bsnm{Sundstrom}, \binits{D.}},
\bauthor{\bsnm{Phillips}, \binits{D.}},
\bauthor{\bsnm{Xie}, \binits{W.}},
\bauthor{\bsnm{Sun}, \binits{Y.}},
\bauthor{\bsnm{Yegitbasi}, \binits{N.}},
\bauthor{\bsnm{Jin}, \binits{H.}},
\bauthor{\bsnm{Hwang}, \binits{E.}},
\bauthor{\bsnm{Shingte}, \binits{N.}},
\bauthor{\bsnm{Berner}, \binits{C.}}:
\bctitle{{Presto: SQL on Everything}}.
In: \bbtitle{{IEEE} {ICDE}},
pp. \bfpage{1802}--\blpage{1813}
(\byear{2019})
\end{bchapter}
\endbibitem

\bibitem[\protect\citeauthoryear{Kim and Lin}{2018}]{Kim:2018}
\begin{bchapter}
\bauthor{\bsnm{Kim}, \binits{Y.}},
\bauthor{\bsnm{Lin}, \binits{J.}}:
\bctitle{{Serverless Data Analytics with Flint}}.
In: \bbtitle{{IEEE} {CLOUD}},
pp. \bfpage{451}--\blpage{455}
(\byear{2018})
\end{bchapter}
\endbibitem

\bibitem[\protect\citeauthoryear{Pu et~al.}{2019}]{Pu:2019}
\begin{bchapter}
\bauthor{\bsnm{Pu}, \binits{Q.}},
\bauthor{\bsnm{Venkataraman}, \binits{S.}},
\bauthor{\bsnm{Stoica}, \binits{I.}}:
\bctitle{{Shuffling, Fast and Slow: Scalable Analytics on Serverless Infrastructure}}.
In: \bbtitle{{USENIX} {NSDI}},
pp. \bfpage{193}--\blpage{206}
(\byear{2019})
\end{bchapter}
\endbibitem

\bibitem[\protect\citeauthoryear{Zhang et~al.}{2021}]{Zhang:2021}
\begin{bchapter}
\bauthor{\bsnm{Zhang}, \binits{H.}},
\bauthor{\bsnm{Tang}, \binits{Y.}},
\bauthor{\bsnm{Khandelwal}, \binits{A.}},
\bauthor{\bsnm{Chen}, \binits{J.}},
\bauthor{\bsnm{Stoica}, \binits{I.}}:
\bctitle{{Caerus: NIMBLE Task Scheduling for Serverless Analytics}}.
In: \bbtitle{{USENIX} {NSDI}},
pp. \bfpage{653}--\blpage{669}
(\byear{2021})
\end{bchapter}
\endbibitem

\bibitem[\protect\citeauthoryear{Congdon}{2018}]{corral}
\begin{botherref}
\oauthor{\bsnm{Congdon}, \binits{B.}}:
{Corral: A Serverless MapReduce Framework Written for AWS Lambda}.
\url{https://github.com/bcongdon/corral/}.
Accessed: 2024-03-26
(2018)
\end{botherref}
\endbibitem

\bibitem[\protect\citeauthoryear{Perron et~al.}{2023}]{Perron:2023}
\begin{barticle}
\bauthor{\bsnm{Perron}, \binits{M.}},
\bauthor{\bsnm{Fernandez}, \binits{R.C.}},
\bauthor{\bsnm{DeWitt}, \binits{D.J.}},
\bauthor{\bsnm{Cafarella}, \binits{M.J.}},
\bauthor{\bsnm{Madden}, \binits{S.}}:
\batitle{{Cackle: Analytical Workload Cost and Performance Stability With Elastic Pools}}.
\bjtitle{Proceedings of the ACM on Management of Data}
\bvolume{1}(\bissue{4}),
\bfpage{233}--\blpage{123325}
(\byear{2023})
\end{barticle}
\endbibitem

\bibitem[\protect\citeauthoryear{Bian et~al.}{2023}]{Bian:2023}
\begin{barticle}
\bauthor{\bsnm{Bian}, \binits{H.}},
\bauthor{\bsnm{Sha}, \binits{T.}},
\bauthor{\bsnm{Ailamaki}, \binits{A.}}:
\batitle{{Using Cloud Functions as Accelerator for Elastic Data Analytics}}.
\bjtitle{Proceedings of the ACM on Management of Data}
\bvolume{1}(\bissue{2}),
\bfpage{161}--\blpage{116127}
(\byear{2023})
\end{barticle}
\endbibitem

\bibitem[\protect\citeauthoryear{Khandelwal et~al.}{2022}]{Khandelwal:2022}
\begin{bchapter}
\bauthor{\bsnm{Khandelwal}, \binits{A.}},
\bauthor{\bsnm{Tang}, \binits{Y.}},
\bauthor{\bsnm{Agarwal}, \binits{R.}},
\bauthor{\bsnm{Akella}, \binits{A.}},
\bauthor{\bsnm{Stoica}, \binits{I.}}:
\bctitle{Jiffy: Elastic far-memory for stateful serverless analytics}.
In: \bbtitle{{ACM} {EuroSys}},
pp. \bfpage{697}--\blpage{713}
(\byear{2022})
\end{bchapter}
\endbibitem

\bibitem[\protect\citeauthoryear{Klimovic et~al.}{2018}]{Klimovic:2018}
\begin{bchapter}
\bauthor{\bsnm{Klimovic}, \binits{A.}},
\bauthor{\bsnm{Wang}, \binits{Y.}},
\bauthor{\bsnm{Stuedi}, \binits{P.}},
\bauthor{\bsnm{Trivedi}, \binits{A.}},
\bauthor{\bsnm{Pfefferle}, \binits{J.}},
\bauthor{\bsnm{Kozyrakis}, \binits{C.}}:
\bctitle{Pocket: Elastic ephemeral storage for serverless analytics}.
In: \bbtitle{{USENIX} {OSDI}},
pp. \bfpage{427}--\blpage{444}
(\byear{2018})
\end{bchapter}
\endbibitem

\bibitem[\protect\citeauthoryear{Wawrzoniak et~al.}{2021}]{Wawrzoniak:2021}
\begin{bchapter}
\bauthor{\bsnm{Wawrzoniak}, \binits{M.}},
\bauthor{\bsnm{M{\"u}ller}, \binits{I.}},
\bauthor{\bsnm{Alonso}, \binits{G.}},
\bauthor{\bsnm{Bruno}, \binits{R.}}:
\bctitle{{Boxer: Data Analytics on Network-enabled Serverless Platforms}}.
In: \bbtitle{CIDR}
(\byear{2021})
\end{bchapter}
\endbibitem

\bibitem[\protect\citeauthoryear{Liao et~al.}{2023}]{Liao:2023}
\begin{botherref}
\oauthor{\bsnm{Liao}, \binits{G.}},
\oauthor{\bsnm{Deshpande}, \binits{A.}},
\oauthor{\bsnm{Abadi}, \binits{D.J.}}:
{Flock: {A} Low-Cost Streaming Query Engine on {FaaS} Platforms}.
CoRR
\textbf{abs/2312.16735}
(2023)
\doiurl{10.48550/ARXIV.2312.16735}
{\href{https://arxiv.org/abs/2312.16735}{{2312.16735}}}
\end{botherref}
\endbibitem

\bibitem[\protect\citeauthoryear{Sharma et~al.}{2016}]{DBLP:conf/eurosys/SharmaGHIS16}
\begin{bchapter}
\bauthor{\bsnm{Sharma}, \binits{P.}},
\bauthor{\bsnm{Guo}, \binits{T.}},
\bauthor{\bsnm{He}, \binits{X.}},
\bauthor{\bsnm{Irwin}, \binits{D.E.}},
\bauthor{\bsnm{Shenoy}, \binits{P.J.}}:
\bctitle{{Flint: Batch-Interactive Data-Intensive Processing on Transient Servers}}.
In: \bbtitle{{ACM} {EuroSys}},
pp. \bfpage{6}--\blpage{1615}
(\byear{2016}).
\doiurl{10.1145/2901318.2901319} .
\burl{https://doi.org/10.1145/2901318.2901319}
\end{bchapter}
\endbibitem

\bibitem[\protect\citeauthoryear{Narasayya and Chaudhuri}{2021}]{DBLP:journals/ftdb/NarasayyaC21}
\begin{barticle}
\bauthor{\bsnm{Narasayya}, \binits{V.R.}},
\bauthor{\bsnm{Chaudhuri}, \binits{S.}}:
\batitle{{Cloud Data Services: Workloads, Architectures and Multi-Tenancy}}.
\bjtitle{Foundations and Trends in Databases}
\bvolume{10}(\bissue{1}),
\bfpage{1}--\blpage{107}
(\byear{2021})
\end{barticle}
\endbibitem

\bibitem[\protect\citeauthoryear{Dageville et~al.}{2016}]{DBLP:conf/sigmod/DagevilleCZAABC16}
\begin{bchapter}
\bauthor{\bsnm{Dageville}, \binits{B.}},
\bauthor{\bsnm{Cruanes}, \binits{T.}},
\bauthor{\bsnm{Zukowski}, \binits{M.}},
\bauthor{\bsnm{Antonov}, \binits{V.}},
\bauthor{\bsnm{Avanes}, \binits{A.}},
\bauthor{\bsnm{Bock}, \binits{J.}},
\bauthor{\bsnm{Claybaugh}, \binits{J.}},
\bauthor{\bsnm{Engovatov}, \binits{D.}},
\bauthor{\bsnm{Hentschel}, \binits{M.}},
\bauthor{\bsnm{Huang}, \binits{J.}},
\bauthor{\bsnm{Lee}, \binits{A.W.}},
\bauthor{\bsnm{Motivala}, \binits{A.}},
\bauthor{\bsnm{Munir}, \binits{A.Q.}},
\bauthor{\bsnm{Pelley}, \binits{S.}},
\bauthor{\bsnm{Povinec}, \binits{P.}},
\bauthor{\bsnm{Rahn}, \binits{G.}},
\bauthor{\bsnm{Triantafyllis}, \binits{S.}},
\bauthor{\bsnm{Unterbrunner}, \binits{P.}}:
\bctitle{{The Snowflake Elastic Data Warehouse}}.
In: \bbtitle{ACM SIGMOD},
pp. \bfpage{215}--\blpage{226}
(\byear{2016})
\end{bchapter}
\endbibitem

\bibitem[\protect\citeauthoryear{Zaharia et~al.}{2021}]{Zaharia:2021}
\begin{bchapter}
\bauthor{\bsnm{Zaharia}, \binits{M.}},
\bauthor{\bsnm{Ghodsi}, \binits{A.}},
\bauthor{\bsnm{Xin}, \binits{R.}},
\bauthor{\bsnm{Armbrust}, \binits{M.}}:
\bctitle{{Lakehouse: {A} New Generation of Open Platforms that Unify Data Warehousing and Advanced Analytics}}.
In: \bbtitle{{CIDR}}
(\byear{2021})
\end{bchapter}
\endbibitem

\end{thebibliography}

\end{document}